\documentstyle[11pt]{article}
\newcounter{subequation}[equation]

\makeatletter 
 
\def\thesubequation{\theequation\@alph\c@subequation}
\def\@subeqnnum{{\rm (\thesubequation)}}
\def\slabel#1{\@bsphack\if@filesw {\let\thepage\relax
   \xdef\@gtempa{\write\@auxout{\string
      \newlabel{#1}{{\thesubequation}{\thepage}}}}}\@gtempa
   \if@nobreak \ifvmode\nobreak\fi\fi\fi\@esphack}
\def\subeqnarray{\stepcounter{equation}
\let\@currentlabel=\theequation\global\c@subequation\@ne
\global\@eqnswtrue
\global\@eqcnt\z@\tabskip\@centering\let\\=\@subeqncr
$$\halign to \displaywidth\bgroup\@eqnsel\hskip\@centering
  $\displaystyle\tabskip\z@{##}$&\global\@eqcnt\@ne
  \hskip 2\arraycolsep \hfil${##}$\hfil
  &\global\@eqcnt\tw@ \hskip 2\arraycolsep
  $\displaystyle\tabskip\z@{##}$\hfil
   \tabskip\@centering&\llap{##}\tabskip\z@\cr}
\def\endsubeqnarray{\@@subeqncr\egroup
                     $$\global\@ignoretrue}
\def\@subeqncr{{\ifnum0=`}\fi\@ifstar{\global\@eqpen\@M
    \@ysubeqncr}{\global\@eqpen\interdisplaylinepenalty \@ysubeqncr}}
\def\@ysubeqncr{\@ifnextchar [{\@xsubeqncr}{\@xsubeqncr[\z@]}}
\def\@xsubeqncr[#1]{\ifnum0=`{\fi}\@@subeqncr
   \noalign{\penalty\@eqpen\vskip\jot\vskip #1\relax}}
\def\@@subeqncr{\let\@tempa\relax
    \ifcase\@eqcnt \def\@tempa{& & &}\or \def\@tempa{& &}
      \else \def\@tempa{&}\fi
     \@tempa \if@eqnsw\@subeqnnum\refstepcounter{subequation}\fi
     \global\@eqnswtrue\global\@eqcnt\z@\cr}
\let\@ssubeqncr=\@subeqncr
\@namedef{subeqnarray*}{\def\@subeqncr{\nonumber\@ssubeqncr}\subeqnarray}
\@namedef{endsubeqnarray*}{\global\advance\c@equation\m@ne%
                           \nonumber\endsubeqnarray}

\makeatletter
\@addtoreset{equation}{section}
\makeatother
\renewcommand{\theequation}{\thesection.\arabic{equation}}

\def\dalemb#1#2{{\vbox{\hrule height .#2pt
        \hbox{\vrule width.#2pt height#1pt \kern#1pt
                \vrule width.#2pt}
        \hrule height.#2pt}}}
\def\square{\mathord{\dalemb{6.8}{7}\hbox{\hskip1pt}}}

\let\a=\alpha \let\b=\beta   \let\e=\epsilon
  \let\q=\theta  
  \let\n=\nu   
     
       \let\D=\Delta

\def\nn{\nonumber} \def\bd{\begin{document}} \def\ed{\end{document}}
\def\ds{\documentstyle} \let\fr=\frac \let\bl=\bigl \let\br=\bigr
\let\Br=\Bigr \let\Bl=\Bigl 
\let\bm=\bibitem
\let\na=\nabla
\let\pa=\partial \let\ov=\overline
\def\ie{{\it i.e.\ }} 
\newcommand{\be}{\begin{equation}} 
\newcommand{\ee}{\end{equation}} 
\def\ba{\begin{array}}
\def\ea{\end{array}}
\def\ft#1#2{{\textstyle{{\scriptstyle #1}\over {\scriptstyle #2}}}}
\def\fft#1#2{{#1 \over #2}}
\def\del{\partial}
\def\sst#1{{\scriptscriptstyle #1}}
\def\oneone{\rlap 1\mkern4mu{\rm l}}
\def\e7{E_{7(+7)}}
\def\td{\tilde}
\def\wtd{\widetilde}
\def\im{{\rm i}}
\def\bog{Bogomol'nyi\ }
\def\q{{\tilde q}}
\def\hast{{\hat\ast}}
\def\0{{\sst{(0)}}}
\def\1{{\sst{(1)}}}
\def\2{{\sst{(2)}}}
\def\3{{\sst{(3)}}}
\def\4{{\sst{(4)}}}
\def\5{{\sst{(5)}}}
\def\6{{\sst{(6)}}}
\def\7{{\sst{(7)}}}
\def\8{{\sst{(8)}}}
\def\n{{\sst{(n)}}}
\def\oo{{\"o}}
\def\hA{\hat{\cal A}}
\def\ns{{\sst {\rm NS}}}
\def\rr{{\sst {\rm RR}}}
\def\tH{{\widetilde H}}
\def\tB{{\widetilde B}}
\def\cA{{\cal A}}
\def\cF{{\cal F}}
\def\tF{{\wtd F}}
\def\Z{\rlap{\sf Z}\mkern3mu{\sf Z}}
\def\ep{{\epsilon}}
\def\IIA{{\rm IIA}}
\def\IIB{{\rm IIB}}
\def\ads{{\rm AdS}}
\def\R{\rlap{\rm I}\mkern3mu{\rm R}}

\def\Ei{{\hbox{Ei}}}
\def\Ci{{\hbox{Ci}}}
\def\Si{{\hbox{Si}}}

\newcommand{\ho}[1]{$\, ^{#1}$}
\newcommand{\hoch}[1]{$\, ^{#1}$}
\newcommand{\bea}{\begin{eqnarray}} 
\newcommand{\eea}{\end{eqnarray}} 
\newcommand{\ra}{\rightarrow}
\newcommand{\lra}{\longrightarrow}
\newcommand{\Lra}{\Leftrightarrow}
\newcommand{\ap}{\alpha^\prime}
\newcommand{\bp}{\tilde \beta^\prime}
\newcommand{\tr}{{\rm tr} }
\newcommand{\Tr}{{\rm Tr} } 
\newcommand{\NP}{Nucl. Phys. }
\newcommand{\tamphys}{\it Center for Theoretical Physics,
Texas A\&M University, College Station, TX 77843}
\newcommand{\upenn}{\it Dept. of Physics and Astronomy, 
University of Pennsylvania,
Philadelphia, PA 19104}

\newcommand{\auth}{M. Cveti\v{c}\hoch{\dagger1}, H. L\"u\hoch{\dagger1}, 
C.N. Pope\hoch{\ddagger2} and T.A Tran\hoch{\ddagger}}

\begin{document}
\begin{flushright}
\hfill{CTP TAMU-1/99 \\ 
UPR/829-T \\
January 1999}\\
\hfill{\bf hep-th/9901002}\\
\end{flushright}


\begin{center}
{\large {\bf Exact Absorption Probability in  the
Extremal Six-Dimensional Dyonic String Background} 
} 

\vspace{20pt}

\auth

\vspace{10pt}
{\hoch{\dagger}\upenn}

\vspace{10pt}
{\hoch{\ddagger}\tamphys}

\vspace{30pt}

\underline{ABSTRACT}
\end{center}

    We show that the minimally coupled massless scalar wave equation
in the background of an six-dimensional extremal dyonic string (or
D1-D5 brane intersection) is exactly solvable, in terms of Mathieu
functions.  Using this fact, we calculate absorption probabilities for
these scalar waves, and present the explicit results for the first few
low energy corrections to the leading-order expressions. For a
specific tuning of the dyonic charges one can reach a domain where the
low energy absorption probability goes to zero with inverse powers of
the logarithm of the energy. This is a dividing domain between the
regime where the low energy absorption probability approaches zero
with positive powers of energy and the regime where the probability is
an oscillatory function of the logarithm of the energy.  By the
conjectured AdS/CFT correspondence, these results shed novel light on
the strongly coupled two-dimensional field theory away from its
infrared conformally invariant fixed point (the strongly coupled
``non-critical'' string).

{\vfill\leftline{}\vfill
\vskip 10pt \footnoterule {\footnotesize \hoch{1} Research supported
in part by DOE grant DOE-FG02-95ER40893
\vskip  -12pt} \vskip   14pt
{\footnotesize
        \hoch{2}        Research supported in part by DOE 
grant DOE-FG03-95ER40917 \vskip -12pt}  \vskip  14pt
}

\pagebreak
\setcounter{page}{1}

\tableofcontents
\addtocontents{toc}{\protect\setcounter{tocdepth}{2}}
\newpage

\section{Introduction\label{sec:intro}}

    One of the more intriguing recent developments in string theory
and M-theory has been the conjecture that when the theory is placed in
a background that is of the form of a product of an $m$-dimensional
anti-de Sitter AdS$_m$ spacetime and an $n$-sphere $S^n$, it is dual
to a superconformal theory on the boundary of the AdS$_m$ spacetime
\cite{mald}.  The AdS$_m\times S^n$ sphere structure arises in the
near-horizon limit of non-dilatonic extremal $p$-branes
\cite{dgt,ght}, and thus one arrives at a conjectured relation between
aspects of the $p$-brane supergravity solution and the superconformal
theory on the AdS$_m$ boundary.

    This conjectured relationship has been explored in a number of
contexts.  One particular aspect has been probed by studying the
absorption probabilities and scattering cross-sections for massless
fields propagating in the supergravity background describing $p$-brane
solitons \cite{gkp,kleb1,gkt}.  In particular, one is principally
interested in studying the absorption probability for incident waves
with low frequency $\omega$.  For example, the leading low-energy
absorption probability for $\ell=0$ partial wave of the the minimally
coupled massless scalar field (dilaton) determines its two-point
correlation function at the conformally invariant infrared fixed point
of the strongly coupled $(m-1)$-dimensional field
theory \cite{gkl,witten,gkpo,fmmr}. Sub-leading corrections are
expected to shed light on the effects of (irrelevant) perturbations
away from the conformally invariant fixed point.

In many of the examples that have been studied, the wave equations
turn out not to be explicitly solvable in closed form, and as a result
various techniques for matching solutions in an overlap between
approximately-solvable inner and outer regions have been used
\cite{unruh,dmw1,gk1,mast,cgkt,km,cl1,cl2,kleb1,gkt,lm,ghkk,mm,taylor,%
dmw2,ah}.  The possibility of making such overlapping approximations
arises because of the low-frequency nature of the incident waves.  One
can then obtain expressions for absorption probabilities as
power-series in $\omega$.  In most cases where such approximations
must be made, it is difficult to obtain more than a few sub-leading
corrections to the leading-order absorption probabilities.

   On the other hand, in certain special cases (of extremal $p$-brane
systems) the wave equations turn out to be exactly solvable, and in
these situations one has more complete control over the calculation of
corrections to the leading-order absorption probability.  One
particularly interesting example arises for the extremal D3-brane,
where the wave equation for massless scalars turns out to be precisely
the (modified) Mathieu equation \cite{gubhash}.  This equation has
been much studied by mathematicians, and techniques have been
developed that are directly adaptable to the problem of calculating
the absorption probability.  Indeed, the work of \cite{doug} formed
the basis of the approach used in \cite{gubhash} for calculating the
D3-brane absorption probability.

   In this paper we consider another example where the wave equation
is exactly solvable, namely the extremal dyonic string \cite{dfkr,dlp}
in six dimensions.  This configuration can alternatively be viewed, by
diagonal dimensional reduction, as being equivalent to an extremal
2-charge black hole in $D=5$, a D1-brane/D5-brane intersection in
$D=10$, or an M2-brane/M5-brane intersection in $D=11$ \cite{tsey}.
From the point of view of the AdS/CFT conjecture this is an especially
interesting case, since the near-horizon geometry of the dyonic string
approaches AdS$_3\times S^3$, and thus the superconformal field theory
\cite{brownh} is defined on the two-dimensional boundary of AdS$_3$.
A controlled method to calculate the absorption probability to an
arbitrary order in energy corrections is thus of interest; such a
result would provide information on the strongly coupled
two-dimensional field theory perturbed away from the infrared
conformally-invariant fixed point by (irrelevant) operators, {\it
i.e.}\ a strongly-coupled ``non-critical string'' would be probed.

    Low energy absorption probabilities for the D1-brane/D5-brane
system, or equivalently the six-dimenesional dyonic string, were
considered in \cite{taylor}, by using the approximate methods
described above, which involve a division of the spacetime into
overlapping inner and outer regions. In particular, the
next-to-leading order energy correction to the s-wave absorption
probability was given there.

In this paper we show that the wave equation for the minimally coupled
massless scalar in the extremal six-dimensional dyonic string
background has in fact an exact solution; a straightforward change of
variables transforms it into the modified Mathieu equation.  Thus we
can again, as was done for the D3-brane \cite{gubhash}, make use of
the results in \cite{doug}, and calculate the absorption probabilities
to an arbitrary order in the energy expansion.  However, there are
some significant differences between the nature of the low energy
expansion of the results presented in this paper and those of the
extremal D3-brane, which we shall explain later in the paper.

The paper is organised as follows.  In section 2, we write down the
scalar wave equation in the background of an extremal dyonic string,
and show that it can be transformed into the Mathieu equation.  In
section 3, we discuss the details of how to solve the Mathieu
equation, paying particular attention to some subtleties associated
with the particular parameter values that arise in the dyonic
background, and their energy dependence.  We obtain the absorption
probability as an energy expansion for a fixed dyonic string
configuration.  In section 4, we explore some different regions in the
parameter space, for which the small expansion parameter is governed
by the ratio of the two dyonic string charges. By tuning the ratio of
the two charges one probes regimes whose absorption probabilities
exhibit novel functional dependence on the energy parameter.

\section{Scalar wave equation for the dyonic string}
\label{II}

   We shall consider a minimally coupled scalar field propagating in
the background of an extremal dyonic string in six dimensions, for
which the metric is
\be
ds_6^2 = (H_e\, H_m)^{-1/2}\, (-dt^2 + dx^2) + (H_e\, H_m)^{1/2} \, 
(dr^2 + r^2\, d\Omega_3^2)\ ,
\ee
where
\be
H_e = 1+ \fft{Q_e}{r^2}\ ,\qquad H_m = 1+ \fft{Q_m}{r^2}\ .
\ee
The wave equation $\square \Phi = 0$ can be separated, by assuming
coordinate dependence
\be
\Phi(t,r,\theta_i) = \phi(r)\, Y(\theta_i)\, e^{-\im\, \omega t}\ ,
\ee
where $Y(\theta_i)$ denotes an harmonic on $S^3$, satisfying $\nabla^2
Y= -\ell(\ell+2)\, Y$ on the unit 3-sphere.  The radial function
$\phi(r)$ therefore satisfies the equation
\be
\fft{d^2\phi}{dr^2} + \fft{3}{r}\, \fft{d\phi}{dr} + 
\Big(\omega^2\, H_e\, H_m -\fft{\ell(\ell+2)}{r^2}\Big) \, \phi = 0\ .
\ee
Defining $\rho=\omega\, r$, the equation becomes
\be
\fft{d^2\phi}{d\rho^2} + \fft{3}{\rho}\, \fft{d\phi}{d\rho} + 
\Big(1 + \fft{\lambda_e^2 +\lambda_m^2-\ell\,(\ell+2)}{\rho^2} +
\fft{\lambda_e^2\, \lambda_m^2}{\rho^4}\Big) \, \phi = 0\ ,\label{waveeq}
\ee
where $\lambda_e^2 =\omega^2\, Q_e$ and $\lambda_m^2 = \omega^2\,
Q_m$.  Note that we would obtain exactly the same equation
(\ref{waveeq}) if we were to consider the intersection of a D1-brane
and a D5-brane in $D=10$, or for such an intersection with NS-NS
branes, or an intersection of an M2-brane and an M5-brane in $D=11$.
We would also encounter the same equation if we considered a 2-charge
black hole in $D=5$.  These equivalences hold because the associated
brane configurations are related to one another by diagonal
dimensional reduction, and the wave equation is invariant under such
reductions on the world volume.

   It is easy to see that the equation (\ref{waveeq}) can be
transformed into the Mathieu equation, by defining 
\be
\phi(\rho) = \fft1{\rho}\, \Psi(\rho)\ ,\qquad 
\rho= \sqrt{\lambda_e\, \lambda_m}\, e^{-z}\ ,
\ee
whence we obtain
\be
\Psi'' + (8\lambda^2 \, \cosh(2z) - 4 \a^2)\, \Psi=0 \ .\label{mathieu}
\ee
Here, we have defined 
\bea
\a^2 &\equiv& \ft14 (\ell+1)^2 - \lambda^2\,\D\ ,\nn\\
\lambda^2 &\equiv& \ft14\lambda_e\, \lambda_m = 
\ft14 \omega^2\, \sqrt{Q_e\, Q_m}\ ,\label{lama}\\
\Delta &\equiv& \fft{\lambda_e}{\lambda_m} +\fft{\lambda_m}{\lambda_e}
=\sqrt{\fft{Q_e}{Q_m}} + \sqrt{\fft{Q_m}{Q_e}} = \fft{Q_e+Q_m}{\sqrt{
Q_e\, Q_m}}\ .\nn
\eea
Note that $\D\ge2$, and that $\lambda$ will be treated as
a small parameter (corresponding to low-energy waves).

     The result that we have found here is in many respects similar to
the one for the D3-brane obtained in \cite{gubhash}, where it was
shown that the scalar wave equation in the D3-brane background could
be reduced to the Mathieu equation.  However, a significant difference
arises in our case, due to the fact that now the parameter $\a$,
appearing in the Mathieu equation, itself depends on the small (energy
dependent) parameter $\lambda$. Although we shall be able to follow
the same general strategy for solving the equation as the one
described in \cite{doug,gubhash}, some adjustments will be required in
order to take into account the additional $\lambda$ (energy)
dependence of parameter $\alpha$.

    The solutions to the Mathieu equation are constructed as an
iterative sequence of terms which correspond to a power series in the
small parameter $\lambda$.  There are various ways in which one can
view the relation between $\a$ and $\lambda$ in equation (\ref{lama}).
The simplest of these is to think of it as defining $\a$ as a power
series in $\lambda$, so that one has $\a=\ft12(\ell+1) + {\cal
O}(\lambda^2)$.  In this viewpoint, one ensures that $\lambda$ is
small by choosing the frequency $\omega$ to be sufficiently small,
with no constraints on the charges. We calculate the absorption
probabilities within this framework in section 3.  One can
alternatively think of $\a$ as being a freely-specifiable parameter
(independent of $\lambda$), where one ensures that $\lambda$, now
viewed as being {\it defined} by the relation between $\a$ and
$\lambda$ in equation (\ref{lama}), is small, by choosing the charges
so that $\D$ is sufficiently large.  We calculate the absorption
probabilities within this framework in section 4.  (In order to
distinguish the somewhat different interpretations of the $\lambda$
parameter in the two viewpoints, we replace $\lambda$ by $\Lambda$ in
section 4 when we are viewing it as a derived quantity.)  It is of
particular interest is to explore the limit where $\alpha\to 0$, since
this approaches the boundary of the region $\alpha^2 \le 0$, which has
a novel low-energy behaviour for the absorption probability.

\section{Solving the Mathieu equation}

    A procedure for solving the Mathieu equation was developed in
\cite{doug}, and formed the basis of the approach used in
\cite{gubhash}.  In order for us to present our notation and results,
and to explain certain subtleties that arise, it will be necessary for
us to review the procedures developed in \cite{doug,gubhash}. At the
same time we shall elaborate on some delicacies in the calculations
that were not explicitly discussed in those papers.

\subsection{The Floquet expansion}

     The Mathieu equation can be solved by mapping the problem into
one of solving a difference equation for coefficients in a series
expansion.  Specifically, the solution can be expressed in the Floquet
form
\be
\Psi(z) = \sum_{n=-\infty}^\infty (-1)^n\, 
C(n+\mu)\, e^{2(n+\mu)\, z}\ , \label{eseries}
\ee
where $\mu$ is a certain constant, related somewhat non-trivially to
the parameters in the Mathieu equation, which will be determined
later.  The quantity $2\mu$ is known as the Floquet exponent, or
Mathieu characteristic exponent.  Substituting (\ref{eseries}) into
(\ref{mathieu}), one finds that the coefficients $C(n+\mu)$ satisfy
the recursion relation
\be
C(x+1) + C(x-1) = \fft{x^2 -\a^2}{\lambda^2}\, C(x)\ .
\label{recursion}
\ee
Convergence of the series (\ref{eseries}) requires that
$C(x+1)<<C(x-1)$, and hence an approximation $C^\0(x)$ to the solution
of this three-term recursion relation can be obtained by solving the
two-term relation
\be
C^\0(x-1) = \fft{x^2 -\a^2}{\lambda^2}\, C^\0(x)\ .
\ee
This is easily seen to lead to
\be
C^\0(x) = \fft{\lambda^{2x}}{\Gamma(x+\a+1)\,
\Gamma(x-\a+1)}\ .\label{c0}
\ee

     A solution to the full recursion relation (\ref{recursion}) can now
be obtained by writing
\be
C(x) = C^\0(x)\, B(x)\ ,\label{cfromb}
\ee
where, substituting into (\ref{recursion}), one finds that the
coefficients $B(x)$ must satisfy
\be
B(x)- B(x+1) = -\fft{\lambda^4}{\big( (x+1)^2-\a^2\big)\, 
\big( (x+2)^2-\a^2\big)}\, B(x+2)\ .\label{brecursion}
\ee
Since $\lambda$ is treated as a small parameter, we can therefore solve
this to any desired accuracy by an iterative procedure, in which the
right-hand side is treated as a ``source term'' constructed using the
solution for the $B(x)$ obtained at the previous iteration.  Thus we
solve
\be
B^{\sst{(i)}}(x) - B^{\sst{(i)}}(x+1) = a(x)\, B^{\sst{(i-1)}}(x+2)\ ,
\ee
where
\be
a(x)\equiv -\fft{\lambda^4}{\big( (x+1)^2-\a^2\big)\, 
\big( (x+2)^2-\a^2\big)}\label{adefn}
\ee
and $B^\0(x) = 1$.  The solution to (\ref{brecursion}) is then given by
\be
B(x) = \sum_{i\ge 0} B^{\sst{(i)}}(x)= 1+ B^\1(x) + B^\2(x)+\cdots\ .
\label{bseries}
\ee
Each successive term in this sum is smaller than its predecessor by a
multiplicative factor of $\lambda^4$, the small expansion parameter of
the iterative solution.\footnote{There is a subtlety concerning this
point, which we shall address in detail later.  For now, we just
remark that the statement is true provided that $B^{\sst{(i)}}(x)$ is
evaluated for a sufficiently large (positive) value of $x$, and that
we can arrange for this criterion to be met.}  It should be emphasised
that in the case we are considering here for the dyonic string, unlike
the D3-brane discussed in \cite{gubhash}, the constant $\a$ in
(\ref{mathieu}) is itself $\lambda$ dependent.  However, this does not
alter the validity of the iterative procedure.

  It is straightforward to see from the above discussion that the
solutions for the quantities $B_n^{\sst{(i)}}$ are given by
\bea
B^\1(x) &=& \sum_{p\ge 0} a(x+p)\ ,\nn\\
B^\2(x) &=& \sum_{p_1\ge 0} \sum_{p_2\ge 2} a(x+p_1)\, a(x+p_1+p_2)\ ,
\nn\\
B^\3(x) &=& \sum_{p_1\ge 0} 
\sum_{p_2\ge 2} \sum_{p_3\ge 2} a(x+p_1)\, a(x+p_1+p_2)\,
a(x+p_1+p_2+p_3)\ ,\\
&\vdots& \nn \\
B^{\sst{(i)}}(x) &=& \sum_{p_1\ge 0} 
\sum_{p_2\ge 2} \cdots\sum_{p_i\ge 2} a(x+p_1)\, a(x+p_1+p_2)\cdots
a(x+p_1+\cdots +p_i)\ .\nn
\eea
Defining 
\be
S(m_0, m_1,\ldots, m_q)\equiv \sum_{i=0}^\infty
\prod_{j=0}^q \big( a(x+i+j)\big)^{m_j}\ , \label{sseries}
\ee
it can then be straightforwardly  shown \cite{doug} that the
coefficients $B^{\sst{(i)}}(x)$ are given by
\be
B^\1(x) = S(1)\ ,\qquad B^\2(x) = \ft12 S(1)^2 -\ft12 S(2) - S(1,1)\ ,
\qquad{\rm  etc}.\ .\label{bfroms}
\ee
The sums $S(m_0, m_1,\ldots, m_q)$ can all be performed explicitly,
with the results expressed in terms of the Digamma and Polygamma
functions.  The expressions for $S(1)$, $S(2)$ and $S(1,1)$ were given
in \cite{gubhash}.  We shall just present the expression for $S(1)$
here, since it is the only sum needed for our present purposes:
\be
S(1) = \fft{(2x+3)\, \lambda^4}{(4\a^2-1)\, (\a^2-(x+1)^2)}
+ \fft{\psi(1+\a+x)-\psi(1-\a+x)}{\a\, (4\a^2-1)}\, \lambda^4\ ,
\label{s1}
\ee
where $\psi(x)\equiv \Gamma'(x)/\Gamma(x)$ is the Digamma function.
(Our definition of $S(m_0, m_1,\ldots, m_q)$ includes a
factor of $(-\lambda^4)^n$ relative to the one in \cite{gubhash},
where $n=\sum_{i=0}^q m_i$ is the rank of $S(m_0, m_1,\ldots, m_q)$.
Thus our $S(1)$ is $(-\lambda^4)$ times the one in \cite{gubhash};
$S(2)$ and $S(1,1)$ have additional factors of $\lambda^8$, and so
on.) As we shall discuss below, the series in (\ref{sseries}) defining
$S(1)$ is convergent, and free of singularities, provided that the
denominators in $a(x)$, given by (\ref{adefn}), do not happen to
vanish for some particular terms in the summation.  In particular,
this means that for generic $x$ the expression (\ref{s1}) should be
finite for all values of the parameter $\a$.  There are apparent
divergences at $\a=\pm\ft12$, but these are actually spurious, and by
taking the limit $\a\rightarrow \pm\ft12$ with due care, the
expression (\ref{s1}) can be seen to be perfectly finite.\footnote{A
similar phenomenon arises also in the expressions that one obtains for
the higher functions $S(m_0,m_1,\ldots,m_q)$.  For example the
expression for the sum $S(1,1)$, given in \cite{gubhash}, has apparent
divergences not only at $\a=\pm\ft12$, but also at $\a=\pm 1$.  Again,
these are spurious, and a careful limiting procedure reveals that
$S(1,1)$ is indeed perfectly finite at these values.}

\subsection{The Floquet exponents}

   The constant $\mu$, related to the Floquet exponent $2\mu$, can be
determined from the observation that $\Psi(z)$ with exponent $-2\mu$ and
$\Psi(-z)$ with exponent $2\mu$ are two Floquet solutions that must be
proportional to one another. From (\ref{eseries}) one then reads off
that the coefficients $C(n+\mu)$ must satisfy the relation
\be
C(n+\mu) = \beta\, C(-n-\mu)\ ,\qquad \hbox{for all}\ \, n\ ,\label{pm}
\ee
where $\beta$ is an $n$-independent constant.  In fact it suffices to
impose (\ref{pm}) for just two consecutive values of $n$, say $n=0$
and $n=1$, since all the rest can then be deduced using
(\ref{recursion}).  Thus the content of (\ref{pm}) is just two
independent equations, which determine the two unknowns $\beta$ and
$\mu$.  In particular, we may conclude that $\mu$ is determined by the
equation
\be
\fft{C(\mu)\, C(-\mu+1)}{C(\mu-1)\, C(-\mu)}= 1\ .\label{mueq}
\ee
The recursion relation (\ref{recursion}) can be re-expressed as
$G(x)=(V(x)-G(x+1))^{-1}$, where $G(x)\equiv
C(x)/C(x-1)$ and $V(x)\equiv (x^2-\a^2)/\lambda^2$.  This may be solved
to give the continued fraction
\be
G(x) = \fft1{V(x)-\fft1{V(x+1)-\fft1{V(x+2)-\cdots}}}\ .
\ee
It follows that (\ref{mueq}), which can be rewritten as
$G(\mu)\,G(1-\mu)=1$, can be solved to give $\mu$ as a power series in
$\lambda$:
\be
\mu = \sum_{n\ge 0} \mu_n\, \lambda^{2n} = \mu_0 + \mu_1\, \lambda^2 
+\mu_2\, \lambda^4 +\cdots\ .
\ee
Note that in our case we obtain a power series in $\lambda^2$, by
contrast to the power series in $\lambda^4$ that arose for the
D3-brane in \cite{gubhash}.  This is because our coefficient $\a^2$ in
the Mathieu equation for the dyonic string is itself $\lambda$
dependent, as given in (\ref{lama}).  For the first four partial
waves, $\ell=0,1,2,3$, we find, up to order $\lambda^8$:
\bea
\ell=0:&&\!\!\! \mu=\ft12 -\sqrt{\D^2-1}\, \lambda^2 -\ft{\D(2\D^2-3)}{
2\sqrt{\D^2-1}}\, \lambda^4 -\ft{(2\D^2-3)(8\D^4-12\D^2+5)}{ 
8(\D^2-1)^{3/2}}\, \lambda^6\nn\\
&&\qquad -\ft{\D(240\D^8-1080\D^6+1798\D^4-1316\D^2+355)}{
48(\D^2-1)^{5/2}}\, \lambda^8 +{\cal O}(\lambda^{10})\ ,\nn\\
&&\nn\\
\ell=1:&&\!\!\! \mu = 1 -\ft12\D\, \lambda^2 -
\ft{3\D^2+8}{24}\, \lambda^4 -
\ft{ 9\D^4 + 88\D^2 -36}{144\D}\, \lambda^6 \nn\\
&&\qquad-
\ft{135\D^6 + 3248\D^4 -2032\D^2+576}{3456\D^2}\, \lambda^8+
{\cal O}(\lambda^{10})\ ,\\
&&\nn\\
\ell=2:&&\!\!\! \mu= \ft32-\ft13\D \, \lambda^2 -\ft{4\D^2+9}{108}\,
\lambda^4 - \ft{\D(16\D^2+117)}{1944}\, \lambda^6 -
\ft{1600\D^4+25380\D^2 +18387}{699840}\, \lambda^8+
{\cal O}(\lambda^{10})\ ,\nn\\
&&\nn\\
\ell=3:&&\!\!\! \mu = 2 -\ft14\D\, \lambda^2 -\ft{15\D^2+32}{960}\,
\lambda^4 -\ft{\D(225\D^2+1504)}{115200}\, \lambda^6 \nn\\
&&\qquad-
\ft{16875\D^4 +235712\D^2+140288}{55296000}\, \lambda^8 +
{\cal O}(\lambda^{10})\ .\nn
\eea
The general formula, up to order $\lambda^8$, turns out to be
\bea
\mu &=& \ft12(\ell+1) -\ft{\Delta}{\ell+1}\, \lambda^2 -
\ft{2+ \ell\, (\ell+2)\, (\D^2+2)}{\ell\, (\ell+1)^3\, (\ell+2)}\,
\lambda^4 -\ft{4(\ell+1)^2\, (3\ell^2 + 6\ell +2)\, \D +2\ell^2\,
(\ell+2)^2\, \D^3}{\ell^2\, (\ell+1)^5\, (\ell+2)^2}\, \lambda^6\nn\\
&&-\ft{\lambda^8}{(\ell-1)\ell^3\,
(\ell+1)^7\, (\ell+2)^3\,
(\ell+3)}\Big\{2(\ell+1)^4(15\ell^4+60\ell^3+
55\ell^2-10\ell-12)\nn\\
&&\qquad\qquad\qquad  +
4(\ell-1)(\ell+1)^2(\ell+3)(15\ell^4+60\ell^3+80\ell^2+40\ell+8)\,
\D^2 
\nn\\
&&\qquad\qquad\qquad+
5(\ell-1)\ell^3\, (\ell+2)^3\, (\ell+3)\, \D^4 \Big\}
+ {\cal O}(\lambda^{10})\ ,\label{mugen}
\eea
when $\ell=2,3,4,\ldots$.  In fact if $\ell$ takes a generic
non-integer value, (\ref{mugen}) is universally valid.  The integer
values for $\ell$ have to be treated carefully because the general
solution for $\mu$ develops poles at the integer values, and one has
to resort to a case-by-case analysis.  It seems that in fact only
$\ell=0$ and $\ell=1$ are not described by the general formula
(\ref{mugen}).\footnote{In this section, we shall present explicit
results for corrections up to and including ${\cal O}(\lambda^6)$ with
respect to the leading-order results for absorption probabilities.
One might think, therefore, that it would be sufficient to determine
$\mu$ up to ${\cal O}(\lambda^6)$, since it has a
$\lambda$-independent constant term for its leading order.  However,
as will be seen in the calculation of the absorption probability $P$
in (\ref{abs1}) and (\ref{abs2}) below, there is a factor of
$|e^{2\im\, \mu\, \pi}- e^{-2\im\, \mu\pi}|^2$ in the expression for
$P$, and so the fact that the constant terms in $\mu$ are integers or
half-integers means that $|e^{2\im\, \mu\, \pi}- e^{-2\im\, \mu\pi}|$
is a quantity of ${\cal O}(\lambda^2)$, rather than the
naively-expected ${\cal O}(1)$.  Consequently, in the calculation of
$|e^{2\im\, \mu\, \pi}- e^{-2\im\, \mu\pi}|$ it is necessary to work
with $\mu$ up to ${\cal O}(\lambda^8)$, for the purposes of eventually
obtaining corrections up to ${\cal O}(\lambda^6)$ in the absorption
probabilities.  Similarly, the occurrence of a pole in
$\Gamma(-\mu-\a+1)$ in (\ref{c0}) means that one must also work with
$\mu$ to order $\lambda^8$ in that expression too, when calculating
$C(-\mu)$ using (\ref{cfromb}). By the same token, in the D3-brane
calculations described in \cite{gubhash}, one should calculate $\mu$
up to ${\cal O}(\lambda^{16})$ in order to obtain corrections up to
${\cal O}(\lambda^{12})$ in the absorption probabilities.}

\subsection{Asymptotic behaviour and absorption probabilities}

    A remarkable property of the Floquet solution of the Mathieu
equation is that one can re-express the series expansion
(\ref{eseries}) in terms of Bessel functions rather than exponentials,
with the {\it same} coefficients $C(n+\mu)$ (up to $(-1)^n$ factors).
(The explanation can be traced back to the fact that exponential
functions and Bessel functions are closely related {\it via} the
Laplace transform \cite{mf}.)  The required solution of the Mathieu
equation, which is in-going on the horizon at $z=\infty$, can be
written in the form
\be
\Psi(z) = {\cal M}(\mu,z)\equiv 
\sum_{n=-\infty}^\infty \fft{C(n+\mu)}{C(\mu)}\,
J_n(2\lambda\, e^{-z})\, H^\1_{n+2\mu}(2\lambda\, e^z)\ ,\label{besseries}
\ee
(The second solution is obtained by choosing an independent Bessel
function, such as $H^\2_{n+2\mu}$, in place of the first Hankel
function $H^\1_{n+2\mu}$.)

    The series (\ref{bseries}) for $\Psi(z)$ is dominated by the $n=0$
term near the horizon, where $z\rightarrow\infty$, and this allows the
asymptotic behaviour in the $z\rightarrow \infty$ region to be read
off easily.  To determine the behaviour at spatial infinity,
corresponding to $z\rightarrow -\infty$, one can make use of the fact,
mentioned earlier, that $\Psi(-z)$ with Floquet exponent $2\mu$ is
proportional to $\Psi(z)$ with Floquet exponent $-2\mu$.  In fact from
(\ref{pm}) one obtains the result that
\be
{\cal M}(-\mu,z)= \fft{C(-\mu)}{C(\mu)}\, {\cal M}(\mu,-z)\ .
\ee
After some further steps described in detail in \cite{gubhash}, in
which one extracts the ratio between the amplitudes of the ingoing and
outgoing waves at spatial infinity, one concludes that the absorption
probability is given by
\be
P= \fft{|\eta -\eta^{-1}|^2}{|\eta -\eta^{-1}|^2 
+|\chi-\chi^{-1}|^2}\ ,\label{abs1}
\ee
where 
\be
\eta= e^{2\im\, \mu\, \pi}\ , \qquad \chi = \fft{C(-\mu)}{C(\mu)}\ .
\label{abs2}
\ee
Note that when determining the orders in $\lambda$ to which one must
work so as to obtain corrections to $P$ to a specified order in
$\lambda$, one must be careful of a number of points.  One such point
was discussed in footnote 3, relating to the fact that the
leading-order terms in $\mu$ are always integers or half-integers.
Another point is that if the leading-order term in $\chi$ is a
$\lambda$-independent constant, then if this constant happens to equal
$\pm1$ then the quantity $|\chi-\chi^{-1}|$, which would naively have
been of ${\cal O}(1)$, would actually instead be of higher order in
$\lambda$.  This in turn would necessitate evaluating $\chi$ to a
higher order in $\lambda$ than one might naively have expected, for
the purposes of finding the corrections to the absorption probability
up to some particular order in $\lambda$.  This particular subtlety
does not arise in our calculations in this section, but it will play a
r\^ole in the calculations in section 4.

     The computation of the absorption probability is now reduced to
the problem of calculating the ratio $C(-\mu)/C(\mu)$, where $C(x)$ is
determined from (\ref{cfromb}), (\ref{bseries}), (\ref{sseries}) and
(\ref{bfroms}).  It is at this point that we meet a subtlety, which
was not explicitly encountered in the general discussion in
\cite{doug}.  It arises because the parameters that occur in our
Mathieu equation are not generic, but are instead rather specific ones
that are fortuitously inconvenient.

     It is evident from (\ref{adefn}) that the sums in (\ref{sseries})
are rapidly convergent, and that the possibility of divergent
behaviour would arise only if the denominators in $a(x)$ were to go to
zero for some particular low term in the summations.  Since $x$, and
$\a$, are functions of the small parameter $\lambda$, the issue that
concerns us here is therefore not that the sums
$S(m_0,m_1,\ldots,m_q)$ might be infinite, but rather that they might
acquire inverse powers of $\lambda$ that could counteract the manifest
$\lambda^4$ numerator in (\ref{adefn}).  If this were to occur, then
our implicit assumption that each term $B^{\sst{(i)}}(x)$ in
(\ref{bseries}) were smaller than its predecessor $B^{\sst{(i-1)}}(x)$
by a factor of $\lambda^4$ would be invalid.  At the very least this
would complicate the discussion because it would mean that we would
have to work to higher order in the iterative solutions
$B^{\sst{(i)}}(x)$ than naively appeared necessary.  At worst, it
could be difficult to determine whether one had included sufficiently
many terms in the iterative solution to give an accurate result to the
desired order in $\lambda$.

   First, let us note that no difficulty of this kind arises in the
calculation of $C(\mu)$.  This is because $\mu= \ft12(\ell+1) + {\cal
O}(\lambda^2)$, and $\a=\ft12(\ell+1) + {\cal O}(\lambda^2)$, and
hence from (\ref{adefn}) it follows that all the quantities
$a(x+i+j)/\lambda^4$ appearing in the sums (\ref{sseries}) are
strictly finite as $\lambda$ tends to zero.  If we define the rank
$n=\sum_{i=0}^q m_i$ for the sum $S(m_0,m_1,\ldots,m_q)$, then it
follows that the quantities $\lambda^{-4n}\, S(m_0,m_1,\ldots,m_q)$,
evaluated at $x=\mu$, are all of the form of a non-zero constant $+
{\cal O}(\lambda^2)$ as $\lambda$ tends to zero, and so the iterative
solutions $B^{\sst{(i)}}(\mu)$ do indeed have the ``naively expected''
form $B^{\sst{(i)}}(\mu) = {\cal O}(\lambda^{4i})$.

    The potential difficulties arise when one considers $C(-\mu)$,
since now we see from (\ref{adefn}) that for sufficiently small values
of the summation variables in (\ref{sseries}) one will encounter
denominators in $a(x)$ that tend to zero as $\lambda$ tends to zero.
This means that these sums $S(m_0,m_1,\ldots,m_q)$, evaluated at
$x=-\mu$, and the associated iterative solutions
$B^{\sst{(i)}}(-\mu)$, will have $\lambda$ dependence whose leading
order is at a more dominant power than the naively expected one.

     The easiest way to handle this problem is to make use once again
of the recursion relation (\ref{brecursion}) for $B(x)$.  By iterating
it $n$ times, we can express $B(-\mu)$ in terms of $B(n-\mu)$ and
$B(n+1-\mu)$.  Thus we can choose $n=\ell+1$, so that the
leading-order constant term in the $\lambda$ expansion of $n-\mu$ is
precisely the same positive constant as occurs in the $\lambda$
expansion of $\mu$ itself.  By this means, we reduce the problem of
calculating $B(-\mu)$, and hence $C(-\mu)$, to the problem of
calculating $B(\ft12(\ell+1)+{\cal O}(\lambda^2))$ and
$B(\ft12(\ell+1)+1+{\cal O}(\lambda^2))$. By our previous discussion,
these cases are easily evaluated, in terms of quantities
$S(m_0,m_1,\ldots,m_q)$ that have precisely the ``expected''
$\lambda^{4k}$ leading-order dependences, where $k=\sum_i m_i$.  Note
that, as mentioned in footnote 3, one must be careful to include one
additional order in $\lambda^2$ in the expansion for $\mu$, over and
above the naively expected one, on account of the pole at $\lambda=0$
in the quantity $\Gamma(-\mu-\a+1)$ that arises in the computation of
$C(-\mu)$ using (\ref{c0}) and (\ref{cfromb}).

    The upshot from the above discussion is that provided we handle
the $C(-\mu)$ calculation carefully, it is necessary to include only
the first iterative correction $B^{\sst{(1)}}(x)=S(1)$ in order to
obtain our results for absorption probabilities with corrections up to
${\cal O}(\lambda^6)$ to the leading-order result.  (By the same
token, the results up to ${\cal O}(\lambda^{12})$ corrections for the
D3-brane, discussed in \cite{gubhash}, require the inclusion only of
the first three iterations $B^{\sst{(1)}}(x)$, $B^{\sst{(2)}}(x)$ and
$B^{\sst{(3)}}(x)$.)  Of course the fact that each term
$B^{\sst{(i)}}(x)$ comes with an overall leading-order $\lambda^{4i}$
factor means that one need calculate the associated sums
$S(m_0,m_1,\ldots,m_q)$ only up to $\lambda^{n-4i}$ corrections in
order to obtain to obtain a results for $B(x)$ to order
$n$. (Assuming, as usual, that the necessary recursion using
(\ref{brecursion}) is performed first, so that the $B^{\sst{(i)}}(x)$
are evaluated at arguments $x\ge\a$.)

   We find that the absorption probability $P_\ell$ for the $\ell$'th
partial wave has the general structure\footnote{The absorption
cross-section $\sigma_{\ell}$ is related to the absorption probability
$P_\ell$ by \cite{gubser}: $\sigma_\ell= {{4\pi}\over
\omega^3}(l+1)^2P_\ell$.}
\be
P_\ell =  \fft{4\pi^2\, \lambda^{4+4\ell}}{(\ell+1)^2 \,
\Gamma(\ell+1)^4}\, \sum_{n\ge0}
\sum_{k=0}^n b_{n,k}\,
\lambda^{2n}\,  (\log \bar\lambda)^k\ ,\label{lamexp}
\ee
where $\bar\lambda = e^\gamma\, \lambda$, and $\gamma$ is Euler's
constant. The prefactor is chosen so that $b_{0,0}=1$.  Our results
for the coefficients $b_{n,k}$ with $k\le n\le 3$, for the first four
partial waves $\ell=0,1,2,3$, are as follows:
\bea
\ell=0:&& b_{1,0} = 4\D\ ,\qquad b_{1,1} = -8\D\ ,\nn\\
&& b_{2,0}= \ft12(32\D^2 -7) -\ft43(\D^2+2)\, \pi^2\ ,\,\,
b_{2,1}= -4(10\D^2+1)\ ,\,\, b_{2,2}= 16(2\D^2 + 1)\ ,\nn\\
&&b_{3,0} =\D\,(64\D^2-31) 
-\ft43\D\, (6\D^2 +17)\pi^2 -\ft{32}{3}\D(\D^2-1)\, 
\zeta(3)\ ,\nn\\
&& b_{3,1}= -8\D\, (22\D^2+1) +\ft{32}{3}\, \D\, (\D^2+5)\, 
\pi^2\ ,\nn\\
&& b_{3,2} = 32\D\, (6\D^2 + 5)\ ,\qquad b_{3,3} = -\ft{256}{3}\D\, 
(\D^2+2)\ ,\nn\\
&&\nn\\
\ell=1:&& b_{1,0} = 5\D\ ,\qquad b_{1,1} = -4\D\ ,\nn\\
&& b_{2,0}= \ft19(144\D^2 +14) -\ft13 \D^2\, \pi^2\ ,\quad
b_{2,1}= -\ft13(63\D^2+8)\ ,\quad b_{2,2}= 8\D^2 \ ,\nn\\
&&b_{3,0} = \ft1{54}\D\, (2241\D^2 + 704) 
-\ft1{18}\D(33\D^2+8)\, \pi^2 -\ft43\D^3\, \zeta(3)\ ,\nn\\
&& b_{3,1}= -\ft1{18}\D\, (1251\D^2+440) +\ft{4}{3}\, \D^3\,  
\pi^2\ ,\nn\\
&& b_{3,2} = \ft43 \D\, (33\D^2 + 8)\ ,\qquad b_{3,3} =
-\ft{32}{3}\D^3\ , \label{6results}\\
&&\nn\\
\ell=2:&& b_{1,0} = \ft{40}{9}\D\ ,\qquad b_{1,1} = -\ft83\D\ ,\nn\\
&& b_{2,0}= \ft1{648}(7472\D^2 +477) -\ft4{27} \D^2\, \pi^2\ ,\,\,
b_{2,1}= -\ft2{27}(164\D^2+9)\ ,\,\, b_{2,2}= \ft{32}{9}\D^2 \ ,\nn\\
&&b_{3,0} = \ft1{5832}\D\, (132992\D^2 + 25443) 
-\ft2{81}\D(28\D^2+3)\, \pi^2 -\ft{32}{81}\D^3\, \zeta(3)\ ,\nn\\
&& b_{3,1}= -\ft2{243}\D\, (3904\D^2+657) +\ft{32}{81}\, \D^3\,  
\pi^2\ ,\nn\\
&& b_{3,2} = \ft{16}{27} \D\, (28\D^2 + 3)\ ,\qquad b_{3,3} =
-\ft{256}{81}\D^3\ ,\nn\\
&&\nn\\
\ell=3:&& b_{1,0} = \ft{47}{12}\D\ ,\qquad b_{1,1} = -2\D\ ,\nn\\
&& b_{2,0}= \ft1{450}(3875\D^2 +171) -\ft1{12} \D^2\, \pi^2\ ,\,\,
b_{2,1}= -\ft1{120}(955\D^2+32)\ ,\,\, b_{2,2}= 2\D^2 \ ,\nn\\
&&b_{3,0} = \ft1{432000}\D\, (6053125\D^2 + 786432) 
-\ft1{1440}\D(485\D^2+32)\, \pi^2 -\ft{1}{6}\D^3\, \zeta(3)\ ,\nn\\
&& b_{3,1}= -\ft2{14400}\D\, (255275\D^2+27488) +\ft{1}{6}\, \D^3\,  
\pi^2\ ,\nn\\
&& b_{3,2} = \ft{1}{60} \D\, (485\D^2 + 32)\ ,\qquad b_{3,3} =
-\ft{4}{3}\D^3\ ,\nn
\eea

    Note that our result for $b_{1,1}$ at $\ell=0$ agrees in
magnitude, but not in sign, with the result obtained in \cite{taylor}.

    For general values of $\ell$ it is quite straightforward to obtain
expressions for the coefficients $b_{1,0}$ and $b_{1,1}$ in
(\ref{lamexp}), using procedures analogous to those that we described
previously.  The only new feature is that in order to calculate
$B(-\mu)$ in terms of $B(x)$ with appropriately positive arguments
$(\ell+1-\mu)$ and $(\ell+2-\mu)$, one must now apply the recursion
relation (\ref{brecursion}) a total of $(\ell+1)$ times.  Up to order
$\lambda^2$, it is easy to see that one need only solve
(\ref{brecursion}) to linear order in $a(x)$, and evaluate
$B(\ell+1-\mu)$ and $B(\ell+2-\mu)$ to zero'th order,
$B(\ell+1-\mu)=1+{\cal O}(\lambda^4)$ and $B(\ell+2-\mu)= 1+{\cal
O}(\lambda^4)$, giving
\be
B(-\mu) = 1 + \sum_{i=0}^{\ell} a(-\mu+i) + \cdots\ .
\ee
Furthermore, it is clear that only those terms for which the denominator
in $a(x)$ in (\ref{adefn}) has zeros at $\lambda=0$, namely $i=\ell$
and $i=\ell-1$, will give ${\cal O}(\lambda^2)$ contributions.  This 
leads to the result
\be
B(-\mu)= 1 +\fft{\lambda^2}{\D\, \ell\, (\ell+2)}+ {\cal O}(\lambda^4)\ .
\ee
After further straightforward calculations, we find that the absorption
probability is given by
\be
P_\ell = \fft{4\pi^2\, \lambda^{4+4\ell}}{(\ell+1)^2\, \Gamma(\ell+1)^4}
\, \Big[ 1 -\fft{8\D}{\ell+1}\, \lambda^2\, \log\lambda +
\fft{4\D\, \lambda^2}{(\ell+1)^2}\, \Big(1+ 2(\ell+1)\, 
\psi(\ell+1)\Big) + \cdots\Big] \ ,
\ee
where $\psi(x)\equiv \Gamma'(x)/\Gamma(x)$ is the Digamma function.
Note that in this case we have not absorbed the Euler gamma constant
$\gamma$ in a rescaling of $\lambda$ in the logarithm.  However, for
all integer $\ell$ the Digamma function has the form $\psi(\ell+1) =R
-\gamma$, where $R$ is a rational number, and so the usual
$\bar\lambda= e^\gamma\, \lambda$ rescaling will enable the Euler
constant to be absorbed in a natural fashion.

\section{Fixed-energy absorption, and ``phase transitions'' }

   So far, we have calculated absorption probabilities in the regime
where $\Delta$ is fixed, and $\lambda$ is a small parameter (see eqn.
(\ref{lama}) for definitions of these quantities).  These results
describe the probabilities as continuous functions of low-energy
incident waves, in a given dyonic string background.  In this section,
we shall study a different regime, where $\alpha$, defined in
(\ref{lama}), is fixed.  Namely, since the background depends on two
charge parameters $Q_e$ and $Q_m$, one can adjust the dyonic
background by tuning $\Delta$ in such a way that $\alpha$ is fixed
(for a fixed value of $\lambda$).  This implies that
\be
\ft14 (\ell+1)^2-\alpha^2=\lambda^2 \, \Delta= \ft14\omega^2\, M
\label{ellalp}
\ee
is fixed, where $M=Q_e+Q_m$ is the mass per unit length of the string. The
low energy expansion  with $\lambda \ll 1$ implies that  $\Delta\gg 1$ (
 $1/\Delta$ can be viewed  as the expansion parameter).  Note that 
in this case $\omega^2\, M$ can be
large, provided that $\D$ is sufficiently large.

     In order not to cause confusion with the previous section, we
write the Mathieu equation as follows:
\be
\Psi'' + (8\Lambda^2\, \cosh(2z) - 4\alpha^2)\, \Psi =0\ ,
\ee
where 
\be
\Lambda^2 \equiv \fft{\ft14(\ell+1)^2
-\a^2}{\D}
\ .\label{lalp}
\ee
In other words $\Lambda$ is now viewed as a function of $\Delta$,
having first specified $\a$.  It is arranged so that $\Lambda$ is 
small by choosing the charges so that $\D$ is sufficiently large.
This should be contrasted with the viewpoint in section 3, where the
parameter $\lambda$ is chosen to be small by taking $\omega$ to be
small, with the charges unconstrained and with $\a$ a derived quantity
following from the expression in equation (\ref{lama}).


   This equation is analogous to the Mathieu equation discussed in
\cite{gubhash}, in that $\a^2$ is now a given fixed parameter,
independent of the expansion parameter $\Lambda$.  However, in our
case $\a^2$ can be arbitrary, whereas in \cite{gubhash} the analogous
parameter could take only discrete integer or half-integer values.  We
see from (\ref{lama}) that depending on the value of $\D$, we can have
$\alpha^2$ either positive, negative or zero.

\subsection{Generic $\alpha^2 > 0$}

    The Floquet exponents in this case turn out to be given by
\bea
\mu &=& \a -\fft{\Lambda^4}{\a(4\a^2-1)}\, - \fft{(60\a^4-35\a^2+2)\,
\Lambda^8}{4\a^3(\a^2-1)(4\a^2-1)^3} \nn\\
&&-\fft{(6720\a^{10} -18480 \a^8 + 15260 \a^6 -4705 \a^4 +413 \a^2
-18)\, \Lambda^{12}}{4\a^5\, (4\a^2-1)^5\, (\a^2-1)^2\, (4\a^2-9)}
+ {\cal O}(\Lambda^{16})\ .\label{mual1}
\eea
This result will be valid for all generic values of $\a$, but breaks
down at the poles occurring when $\a$ is an integer or half-integer.
We shall return to a discussion of these special cases later, and
concentrate for now on the generic situation when (\ref{mual1}) is
valid.  It is useful to define
\be
\a= \ft12 L + 1\ ,
\ee
so that for now, we shall consider cases where $L$ is not an integer.
For such generic values of the parameter $\alpha$ there are no
subtleties involving the occurrence of poles or zeros in the
calculation of $C(-\mu)$, nor in the expression $|\eta-\eta^{-1}|$
appearing in (\ref{abs1}).  We shall just present results for the
first two corrections to the leading-order absorption.  One has to
distinguish between two cases, according to whether $\a>\ft12$ or
$\a<\ft12$ (we shall consider the case $\a=\ft12$ later).  The reason
for this distinction is that when $\a>\ft12$ the next-to-leading order
corrections are always of the form $\Lambda^4$ and $\Lambda^4\,
\log\Lambda$, whilst if $\a<\ft12$ the next-to-leading order
correction is of the form $\Lambda^{8\a}$.  For $\a>\ft12$, we find
that the absorption probability is given by
\be
P_\ell(\a)= \fft{\pi^2\, \Lambda^{8\a}}{\a^2 \Big[ \Gamma(-2\a)^2 -
\Gamma(2\a)^2\Big]}\, \Big(1 + b_{1,1}\, \Lambda^4 \, \log\Lambda +
b_{1,0}\, \Lambda^4 + \cdots \Big)\ ,\label{alprob1}
\ee
where
\bea
b_{1,1} &=& \fft{8 \Big[\Gamma(-2\a)^2 + \Gamma(2\a)^2\Big]}{
\a\, (4\a^2-1)\,\Big[\Gamma(-2\a)^2 - \Gamma(2\a)^2\Big]} \ ,\nn\\
b_{1,0} &=& -\fft{4\pi\cot 2\pi\a}{\a\, (4\a^2-1)} + \ft12 b_{1,1}\, 
\Big( \fft{4\a^2}{4\a^2-1} - \psi(-2\a) -\psi(2\a)\Big)\ ,
\eea
where $\psi$ is again the Digamma function.

   When $\a<\ft12$, we find that up to the first correction to the
leading order, the absorption probability is
\be
P_\ell(\a) =  \fft{4\Lambda^{8\a}\, \Gamma(-2\a)^2 \sin^22\pi\a}{
\Gamma(2\a)^2}\, \Big( 1 + 2\Lambda^{8\a}\, \fft{\Gamma(-2\a)^2}{
\Gamma(2\a)}\, \cos 4\pi\a + \cdots \Big)\ .\label{alprob2}
\ee
The next correction will be either $\Lambda^{16\a}$, or $\Lambda^4\,
\log\Lambda$, depending on whether $\a<\ft14$ or $\a\ge\ft14$. 

\subsection{$\a=\ft12L + 1$}

    In the previous subsection, we saw that for generic values of
$\a\equiv \ft12 L+1$, corresponding to $L$ non-integer, the
leading-order absorption probability is of the form
\be
P_{\sst L} \sim \Lambda^{4L}\ .
\ee
This result is not applicable if $L$ is an integer.  This can be seen
both from the resulting poles in the expression (\ref{mual1}) for
$\mu$, and the singular behaviour of $\Gamma(-2\a)$ and the
trigonometric functions in (\ref{alprob1}) and (\ref{alprob2}).  In
fact the absorption probabilities for the special cases $L=$ integer
are precisely those calculated in \cite{gubhash}.  These have the
leading-order form
\be
P_{\sst L} \sim \Lambda^{8+4L}\ .
\ee
To be precise, the absorption probabilities in these cases are given by
\cite{gubhash}
\be
P_{\sst{L}} = \fft{4\pi^2\, \Lambda^{8+4L}}{(L+1)^2\, \Gamma(L+1)^4} \, 
\sum_{n\ge0}\sum_{k=0}^n b_{n,k}\, \Lambda^{4n}\, (\log\bar\Lambda)^k
\ ,
\ee
where $\bar\Lambda= e^\gamma\, \Lambda$.  The explicit results for the
coefficients $b_{n,k}$ with $k\le n\le 3$ were given for $L=0,1,2$ in
\cite{gubhash}.  In order to calculate the corrections up to this
$\Lambda^{12}$ order, it is necessary to calculate the Floquet
exponent $2\mu$ up to order $\Lambda^{16}$, for the reasons that we
discussed in footnote 3.  They are given by
\bea
L=0:&& \mu= 1-\ft{\im\, \sqrt5}{6}\, \Lambda^4 + \ft{7\,\im}{216\sqrt5}\,
\Lambda^8 + \ft{11851\,\im}{62208\sqrt5}\, \Lambda^{12} -
\ft{1243157\, \im}{ 55987200\sqrt5}\, \Lambda^{16} + {\cal O}(\Lambda^{20})\
,\nn\\ L=1:&& \mu =\ft32 -\ft1{12} \Lambda^4 + \ft{133}{8640}\,
\Lambda^8 +
\ft{311}{3110400}\, \Lambda^{12} + \ft{908339}{1254113280}\,
\Lambda^{16} + {\cal O}(\Lambda^{20})\ ,\\
L=2:&& \mu=2 -\ft1{30}\, \Lambda^4 - \ft{137}{54000}\, \Lambda^8 +
\ft{305843}{1360800000}\, \Lambda^{12} - \ft{64347197}{
1714608000000}\, \Lambda^{16} + {\cal O}(\Lambda^{20})\ .\nn
\eea
It is also necessary to calculate the sums $S(m_0,m_1,\ldots,m_q)$ up to
rank $n\equiv\sum_i m_i=3$.  The expressions for $n=1$ and $n=2$ are
given in \cite{gubhash}.  

There are in fact special values of the integer $L$ which themselves
must be treated as special cases.  One of these is $L=-2$,
corresponding to $\a^2=0$. The other is $L=-1$, corresponding to
$\a^2=\ft14$. These two cases are discussed in the next two
subsections.

\subsection{$\alpha^2=\ft14$}

   In this case, the Floquet exponent $2\mu$ is given by
\be
\mu = \ft12 + \im\,  \Lambda^2 -\ft{15\, \im}{8}\, \Lambda^6 +
+\ft{9707\, \im}{1152}\, \Lambda^{10} +
{\cal O}(\Lambda^{14})\ .
\ee
We find that the absorption probability is then given by
\be
P=4\pi\, \Lambda^4\, \Big(1 + 16\Lambda^4\, (\log \Lambda)^2 -
4\Lambda^4\, \log\Lambda -(\ft72 +\ft83\, \pi^2)\, \Lambda^4 +
\cdots \Big)\ .
\ee
In fact this result can also be read off directly from the $\ell=0$
expressions in (\ref{6results}), by setting $\D=0$.

\subsection{$\alpha^2=0$}

     When $\a=0$, we find that the Floquet exponent is
given by
\be
\mu=\sqrt2 \Lambda^2\, ( 1 + \ft{25}{8} \Lambda^4 +
\fft{31783}{1152} \Lambda^8 
+\fft{26498155}{82944} \Lambda^{12}
+ {\cal O}(\Lambda^{12}))\ .\nn\\
\ee
We then obtain the absorption probability 
\be
P= \fft{\pi^2}{\pi^2+ (2\log\bar\Lambda)^2}\, 
\Big(1 -\ft{32}{3}\Lambda^4\, (\log\bar\Lambda)^2
-\ft{16}{3}\, (4\zeta(3)-3) \, \fft{\Lambda^4\, \log\bar\Lambda}{
\pi^2+ (2\log\bar\Lambda)^2}
+{\cal O}(\Lambda^8)
\Big)\ ,\label{a0}
\ee
where $\bar\Lambda= e^\gamma\, \Lambda$.  Note that as $\Lambda\to 0$
the absorption probability (\ref{a0}) approaches zero not as a
positive (integer) power of $\Lambda$, but instead as $(\log{\bar
\Lambda})^{-2}$. This result indicates that on the field theory side,
the infrared limit with the specific choice of the dyonic background
that ensures $\alpha=0$ exhibits new features.  For example, the
two-point correlation function for the minimally-coupled
scalar~\cite{gkl} has a leading non-zero contribution that is
inherently quantum in nature (as signalled by the $\log{\bar\Lambda}$
dependence).  Note that the $\a^2=0$ case that we have considered here
is an example of the kind we mentioned in section 3.2, where the
$\Lambda$-independent term in the expansion for $\chi$ is equal to 1,
implying that one loses the naively-expected leading-order constant
when one evaluates $|\chi-\chi^{-1}|$.

It would be interesting to explore the effects of infinitesimal
deviations from the limit $\alpha=0$.  It turns out that if one takes
a special slice $\alpha^2=b^2\Lambda^4$, where $b=$ constant, one
obtains a systematic expansion of $\mu$ around small $\Lambda$:
\be
 \mu=\sqrt{2}\,\Lambda^2\, \Big(\sqrt{1+{b^2\over 2}}+{{{25\over
8}+2b^2}\over \sqrt{1+{b^2\over 2}}}\Lambda^4+{\cal O}(\Lambda^8)
\Big)\ .
\ee
In this  case  the absorption probability has the  following form:
\be
P= \fft{\pi^2}{\pi^2+ (2\log\bar\Lambda)^2}\, 
\Big(1 -\ft{8}{3}(1+\ft{b^2}{2})\Lambda^4\, (2\log\bar\Lambda)^2
-\ft{8}{3}\, [\zeta(3)(8+b^2)-6] \, \fft{\Lambda^4\, \log\bar\Lambda}{
\pi^2+ (2\log\bar\Lambda)^2}
+{\cal O}(\Lambda^8)
\Big)\ ,\label{a00}
\ee

Note that as one approaches $\alpha^2\sim b^2\Lambda^4$, only the
sub-leading term in the $\Lambda$ expansion depends on $b^2$.

\subsection{$\alpha^2<0$}

   One can see from (\ref{lalp}) that it is possible to choose the
parameters so that $\a^2$ becomes negative, while keeping $\Lambda$
small.  In such a case, we may write
\be
\a=\im\, \beta\ ,
\ee
where $\beta$ is real, and related to $\Delta$ by
\be
\Lambda^2 = \fft{\ft14(\ell+1)^2 +\b^2}{\Delta}\ .
\ee
In this case the Floquet exponents are given by (\ref{mual1}) with
$\a=\im\, \b$.  The absorption probability is now qualitatively
different in form.  In particular, the leading-order contribution
becomes oscillatory as a function of $\Lambda$; we find that it is given
by
\be
P= \fft{\sinh^2 2\pi\b}{\sinh^2 2\pi\b + \sin^2(\theta-4\b\,
\log\Lambda)} +\cdots \ ,
\ee
where 
\be
\theta={\rm arg}\, \fft{\Gamma(2\im\, \b)}{\Gamma(-2\im\, \b)}\ .
\ee
The oscillations become insignificant if $\beta>>1$. Again this region
exhibits novel functional dependence of the absorpiton probability on
$\Lambda$, thus signalling novel field theoretical phenomena on the
dual side.

\section{Conclusions}

    In this paper, we have studied the solutions of the minimally
coupled scalar wave equation in the background of a six-dimensional
extremal dyonic string. Owing to the fact that the equation is
equivalent to the modified Mathieu equation, the wave-functions can be
constructed exactly, in the sense that one can obtain exact solutions
as power series in the parameter $\lambda^2$ appearing in the Mathieu
equation (\ref{mathieu}).  This provides a systematic power-series
expansion that is quite different in nature from a power-series
solution of a differential equation in some restricted range of the
independent variable.  In particular, it allows one to have complete
control over the relation between the small-distance and
large-distance asymptotics of the wave functions; it is this
information that is needed in order to calculate the absorption
probabilities.

     Since the dyonic string background depends on two charge
parameters $Q_e$ and $Q_m$, there are two independent dimensionless
quantities that can play the r\^ole of ordering parameters in the
iterative solution of the Mathieu equation.  One of these is the
quantity $\lambda=\ft14\omega^2\,\sqrt{Q_e\, Q_m}$, which can be
arranged to be small by taking the frequency $\omega$ of the wave
sufficiently small.  The other parameter is
$1/\D=\sqrt{Q_eQ_m}/(Q_e+Q_m)\le 1/2$, which provides a measure of the
ratio between the electric and magnetic charges $Q_e$ and $Q_m$.  In
section 3 we calculated the absorption probability at low frequency,
with $\D$ fixed, {\it i.e.}\ with a fixed dyonic string background.
In this case the leading-order constant term in the expansion for $\a$
can take only discrete values, namely the integers and half-integers
$\ft12(\ell+1)$, where $\ell$ characterises the angular dependence of
the $\ell$'th partial wave.  The low-frequency expansion arises in
even powers of $\lambda$ along with the characteristic dependence on
powers of $\log{\bar \lambda} $ (${\bar \lambda}=e^\gamma\lambda$,
where $\gamma$ is Euler's constant). These results are analogous in
structure to those for other extremal black-hole or $p$-brane
backgrounds.  For example, the occurrence of $\log{\bar\lambda}$ in
sub-leading terms in the energy expansion for the absorption
probability is analogous to what is seen in other examples, such as
the M-branes and D3-brane \cite{ghkk,gubhash}. By the AdS/CFT
correspondence, the results of section 3 should yield information on
correlation functions in strongly coupled two-dimensional quantum
field theory perturbed away from the infrared conformally invariant
fixed point.
     
     In section 4 we calculated absorption probabilities in a
different parameter regime, namely where $\a$ is held fixed, with
$1/\D$ being used as the small order-parameter of the iterative
solution of the Mathieu equation.  This can be viewed as an expansion
in which an incident wave of fixed frequency (satisfying $\lambda\ll
1$) is absorbed in the dyonic string background with a range of charge
ratios, in a limit where $\Delta$ is sufficiently large, implying that
the charge ratio is very large, In this case the value of $\a$ is
fixed, but can be chosen to have an arbitrary value.  We first
obtained results for generic positive values of $\a^2$.  The
expressions for Floquet exponents and absorption probabilities become
singular when $\a$ takes integer or half-integer values, and these
cases have to be treated separately; this was discussed in section
4.2.

   A case of particular interest which itself lies outside the general
discussion of special cases in section 4.2 is when $\a=0$, discussed
in section 4.4.  In this situation, we find that the absorption
probabilities depend inversely on powers of $\log\Lambda$, which seems
to indicate a correspondence to an intrinsically quantum regime of the
two-dimensional field theory.  The $\a=0$ point can be viewed as a
``critical point'', since the absorption probabilities for $\a^2>0$
and for $\a^2<0$ (discussed in section 4.5) are qualitatively
different.  When $\a^2>0$, the absorption probability vanishes in the
limit $\lambda\longrightarrow 0$ and approaches this limit with a
positive (multiple of 4) power of $\lambda$.  On the other hand when
$\a^2<0$, the leading-order behaviour becomes an oscillatory function
of $\log\lambda$.  We also studied the region close to the critical
point $\alpha=0$, by taking $\a^2 = b^2\, \lambda^4$, with $b$
constant.

    Owing to the fact that the boundary conformal field theory for the
extremal dyonic string is two-dimensional, this example provides an
especially manageable arena for making concrete tests of the AdS/CFT
conjecture.  The fact that on the supergravity side one has full
control over the study of the scattering problem provides a starting
point for addressing the corresponding strongly coupled non-critical
two-dimensional field theory.

\section*{Acknowledgements} 

     After this work was completed,  we learned from Steven Gubser that
he and Akikazu Hashimoto  were also aware that the wave equation in the
extremal dyonic string  background could be transformed into the Mathieu
equation \cite{GH}. We are grateful to him for  correspondence on his
unpublished work \cite{GH} and useful comments. We also acknowledge
helpful discussions with Steve Fulling.  We have made extensive use of
Mathematica and Maple for calculations.

\end{document}